\documentclass[structabstract]{aa}

\usepackage{natbib,twoopt}
 \usepackage[breaklinks=true]{hyperref} 
 \bibpunct{(}{)}{;}{a}{}{,}    
 \newcommandtwoopt{\citeads}[3][][]{\href{http://adsabs.harvard.edu/abs/#3}%
                                        {\citealp[#1][#2]{#3}}}
 \newcommandtwoopt{\citepads}[3][][]{\href{http://adsabs.harvard.edu/abs/#3}%
                                        {\citep[#1][#2]{#3}}}
 \newcommandtwoopt{\citetads}[3][][]{\href{http://adsabs.harvard.edu/abs/#3}%
                                        {\citet[#1][#2]{#3}}}
 \newcommandtwoopt{\citeyearads}[3][][]%
   {\href{http://adsabs.harvard.edu/abs/#3}{\citeyear[#1][#2]{#3}}}

\usepackage{amssymb}
\usepackage{graphicx}
\usepackage{longtable,lscape,supertabular}
\usepackage{txfonts}
\usepackage{natbib}

\usepackage{times,mathptm}

\usepackage{epsfig}
\def\ergs{erg s$^{-1}$}
\def\ergcmsec{erg cm$^{-2}$ s$^{-1}$}

\begin{document}

\title{Seyfert's Sextet: where is the gas?}

\author{S. Tamburri\inst{1}
\and
G. Trinchieri\inst{1}
\and
A. Wolter\inst{1}
\and
J. Sulentic\inst{2}
\and
A. Durbala\inst{3}
\and
M. Rosado\inst{4}
}
\institute{ 
INAF--Osservatorio Astronomico di Brera, Via Brera 28, 20121 Milano, Italy
\\email: {\texttt{sonia.tamburri@brera.inaf.it}}
\and 
Instituto de Astrofisica de Andalucia, 18008 Granada, Spain
\and
University of Wisconsin -Stevens Point, Department of Physics and Astronomy, 2001 Fourth Avenue, Stevens Point, WI 54481-1957
\and
Instituto de Astronomia, Universidad Nacional Autonoma de Mexico (UNAM), Apdo. Postal 70-264, 04510, Mexico, D.F., Mexico
}


\date{Received 29 December 2011 / Accepted 13 march 2012}

\abstract{}{Seyfert's Sextet (a.k.a HCG 79) is one of the most compact and isolated galaxy groups in the local Universe. It shows a prominent diffuse light component that accounts for $\sim 50$\% of the total observed light. This likely indicates that the group is in an advanced evolutionary phase, which would predict  a significant hot gaseous component. Previous X-ray observations had suggested a low luminosity for this system, but with large uncertainties and poor resolution.}
{We present the results from a deep (70 ks), high resolution \textit{Chandra} observation of Seyfert's Sextet, requested with the aim of separating the X-ray emission associated with the individual galaxies from that of a more extended inter-galactic component. We discuss the spatial and spectral characteristics of this group we derive with those of a few similar systems also studied in the X-ray band.} {The high resolution X-ray image indicates that the majority of the detected emission does not arise in the compact group but is concentrated towards the NW and corresponds to what appears to be a background galaxy cluster. The emission from the group alone has a total luminosity of $\sim$ 1$\times 10^{40}$ \ergs\  in the (0.5-5) keV band.  Most of the luminosity can be attributed to the individual sources in the galaxies, and only $\sim$ 2$\times 10^{39}$ \ergs\ \  is due to a gaseous component. However, we find that this component is also mostly associated with the individual galaxies of the Sextet, leaving little or no residual in a truly IGM component. The extremely low luminosity of the diffuse emission in Seyfert's Sextet might be related to its small total mass.}{}

\maketitle

\authorrunning{S. Tamburri et al.}
\titlerunning{SS: where is the gas?}

\section{Introduction}

Seyfert's Sextet (a.k.a \object{HCG 79}) is one of the most compact and isolated galaxy aggregates in the local Universe\footnote{$v_r\sim4400\ \rm{km\ s^{-1}}$ (\citealt{durbala}) implies $D\sim60\ \rm{Mpc}$ for H$_0=73\ \rm{km\ s^{-1} Mpc^{-1}}$. At the distance of the group 1' is equivalent to 17.1 kpc} (\citealt{hickson82}, \citealt{sulentic87}, \citealt{iovino}). 
The group involves four accordant redshift galaxies (H79abcd) plus what we interpret as  
a largely dissolved remnant galaxy H79f, which some would call a tidal tail (see Fig. \ref{fig:1}-left for the optical image and labeling). A background galaxy H79e appears projected within the boundaries of the group ($v_r\sim19809\ \rm{km\ s^{-1}}$). 
Basic properties of the galaxies are listed in Table \ref{tab:1}.
In the context of evolution of galaxy aggregates, a group of early-type galaxies embedded in a stellar halo must represent an evolved system while groups containing mostly late-type galaxies should be in earlier evolutionary phases (\citealt{Mulchaey}). 
Multiwavelength observations of HCG 79 support the hypothesis that it is a highly evolved system probably still growing by occasional infall of late-type neighbors. This must be a very slow process because HCG 79 has few neighbors to accrete within several Mpc (\citealt{durbala}, D08 hereafter).
The extreme compactness of Seyfert's Sextet (the group diameter is $\sim 1.3$ arcmin, $\sim 23$ kpc), 
the low velocity dispersion ($\sigma_v\leq 200$ km/s), the number of early-type members (3 out of 5, Table \ref{tab:1}), the fraction of total light in a diffuse optical component ($\sim 40-50\%$), the star formation suppression and the lack of a high luminosity AGN suggest that it should be at a late evolutionary stage. 
Some of these features are also observed in other evolved compact groups, which show an X-ray luminosity L$_X\gtrsim5\times10^{41}$ \ergs\  (e.g. HCG 15, 51, 42, 62, 92; \citealt{rasmussen}, \citealt{sun}, \citealt{finoguenov}, \citealt{morita}, \citealt{trinchieri03,trinchieri05}). Another clue that HCG 79 is relatively old and evolved comes from the sizes and luminosities of its early-type members. They are too bright for their sizes, consistent with stripped spiral bulges. This was demonstrated by comparing sizes and luminosities of HCG 79 members with the bulges of typical spirals in the HCG 79 neighborhood (D08). 
In the context of galaxy structure evolution, groups are expected to become more prominent in the X-ray band as they grow older, therefore in this scenario it would be reasonable to expect a strong diffuse and relaxed X-ray component associated with HCG 79.

VLA maps (\citealt{verdes}) suggested that HCG 79 is gas deficient in HI, although the expected HI content itself is somewhat uncertain, due to the uncertainties on the morphology of its original members. Subsequently, however, single-dish GBT observations detected an extended, very low surface brightness HI component (\citealt{borth}) missed by the higher resolution VLA map, better reconciling the amount of neutral HI gas with expectations.
At the same time, a search for additional gas, hidden in a hot component, motivated us to ask for deep \textit{Chandra} observations of this system, which we are now presenting here.

\begin{table}
\caption{Properties of the member galaxies of Seyfert's Sextet and discordant galaxy H79e.} 
\label{tab:1}
\centering
\begin{tabular}{cccccc}\hline
\small{HCG 79} & v$_r$ & Type & D$_{r}$ & L$_{r}$ & L$_{X}$\\
\small{ } & (km s$^{-1}$) &  & (kpc)&\ ($10^9$ L$_\odot$) & \ergs\ \\ \hline 
a & 4292 & E3 & 8.1 & 7.9 &$2.6\times10^{39}$\\ 
b & 4446 & S0 & 6.5 & 8.7 &$1.9\times10^{39}$\\ 
c & 4146 & S0 & 5.3 & 3.5 & $< 1\times10^{39}$\\
d & 4503 & Sd & 7.6 & 1.4 &$3.5\times10^{39}$\\
e & 19809 & Sc & -- & -- & --\\
f & 4095 & remnant & 5.1 & 0.7 & --\\\hline
\end{tabular} 
\begin{list}{}{}
 \item[ ]{\textbf{Notes.} $D_r$ is the semimajor axis of the last concentric isophote in the \textit{r} band. Data taken from \cite{durbala}. L$_{X}$ in the (0.5-5) keV energy band from this paper, see section 2.4.}
\end{list}
\end{table}

\cite{pildis} reported an X-ray detection of HCG 79 at 2.6 $\sigma$ level with ROSAT/PSPC in the soft energy band (0.1-2.4) keV. The observation yielded only $28.4\pm11.1$ net counts which, considering the relatively poor PSPC resolution, could not be properly attributed to the different components of HCG 79. \cite{nishiura} subsequently used the ROSAT data and estimated a soft X-ray luminosity of L$_X\sim2-3\times10^{40}$ \ergs\  for the group.
They further suggested that, while the ROSAT data did not resolve the emission, it was nonetheless indicative of the presence of a soft X-ray halo morphologically similar to the diffuse optical light.

In this work, we present a detailed analysis of a pointed \textit{Chandra} observation of HCG 79. The observation was requested in order to separate the individual galaxy contribution from that of the diffuse group emission, and to study the X-ray morphology and spectral characteristics of the different components. We discuss Seyfert's Sextet in the context of other compact groups and their evolution. 
The structure of the paper is as follows: in Section 2 we present the X-ray data of Seyfert's Sextet obtained with \textit{Chandra}, in Section 3 we describe the main properties of the background cluster we detect with this data, in Section 4 we summarize our results.

\section{Data Analysis and Results}
HCG 79 was observed with \textit{Chandra}'s back-illuminated CCD S3 (ACIS-S in imaging configuration) on May 20, 2010 for a total of 70 ks, observation ID 11261. The basic X-ray data reduction was based on the \textquotedblleft CIAO science threads\textquotedblright\  given on the \textit{Chandra} X-Ray Center (CXC) Web site\footnote{{http://cxc.harvard.edu/ciao/}}. 
We also used \textquotedblleft Sherpa\textquotedblright\ \footnote{http://cxc.harvard.edu/sherpa/index.html}, the \textquotedblleft funtools\textquotedblright\ tasks, ds9 and corollary software\footnote{{http://hea-www.harvard.edu/RD/index.html}}. For the spectral analysis we used XSPEC\footnote{{http://heasarc.nasa.gov/xanadu/xspec/}} 12.6.0.
No background ``flares'' affected the observations, therefore the useful exposure time is 70.0 ks.
All quantities are derived from the event file in the broad energy band (0.5-5.0) keV chosen to maximize the signal.

\subsection{X-ray Images}

\begin{figure*}
\resizebox{\hsize}{!}{\includegraphics{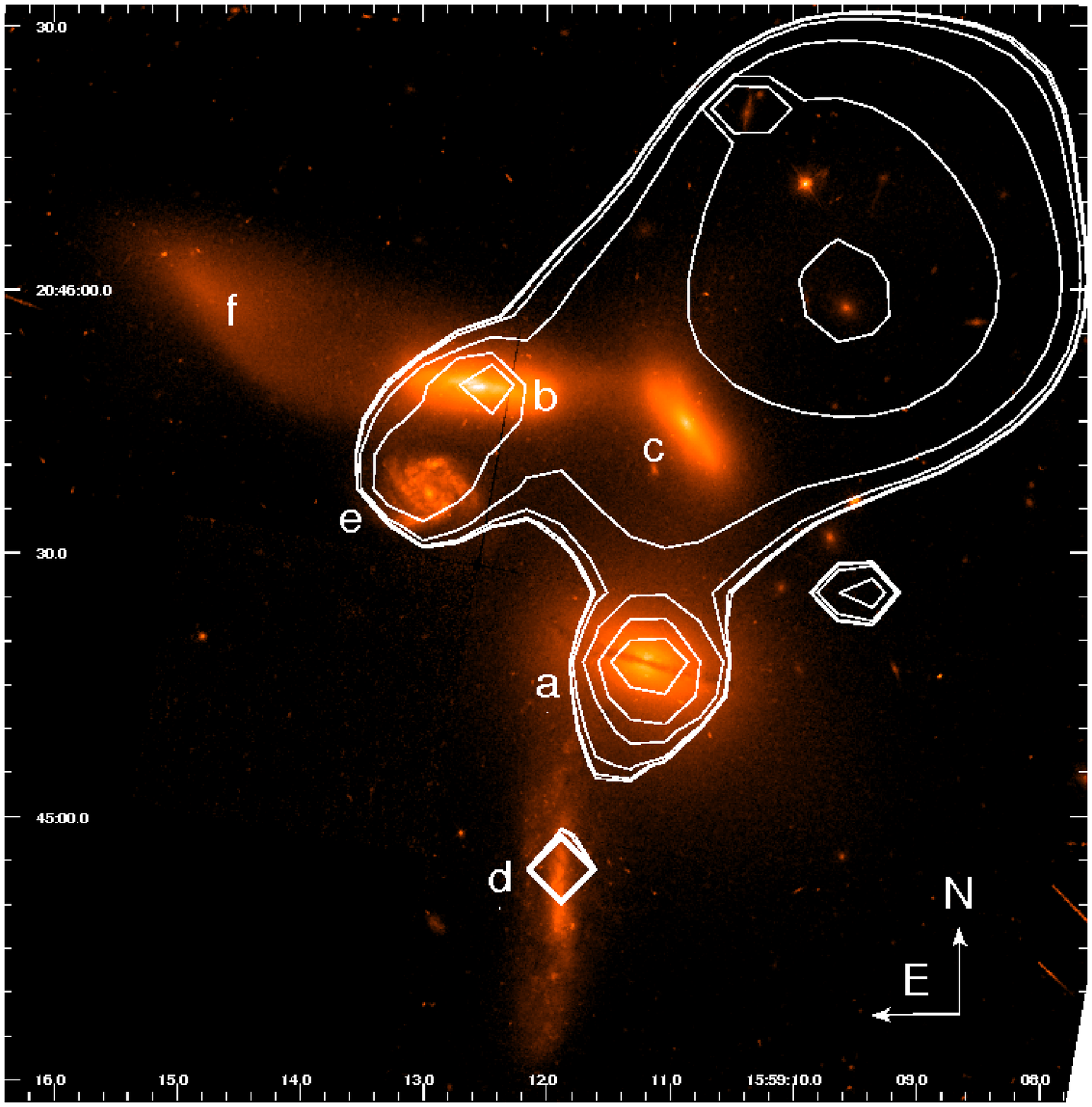}
\includegraphics{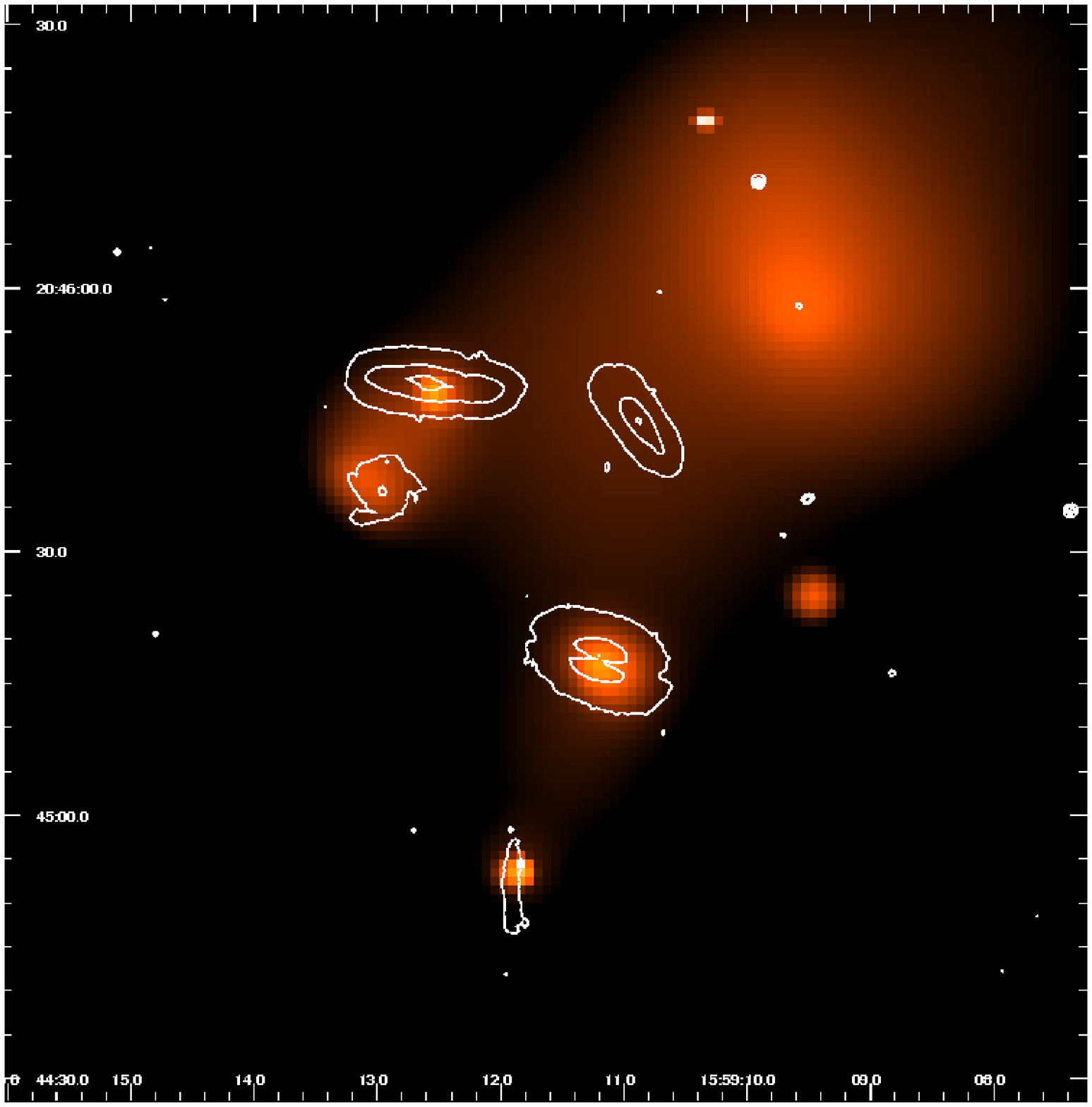}}
\caption{\textit{Left}: optical image of the Seyfert's Sextet from the HST archival image [WFPC2, F555W filter] with (0.5-5) keV X-ray contours superimposed; the contours are logarithmically spaced and the first displayed is at $0.29 \times 10^{-5}$ ct s$^{-1}$ pix$^{-1}$. The original Hickson labeling of the galaxies is indicated. \textit{Right}: raw smoothed \textit{Chandra} image with optical contours of HCG 79 and brighter sources overlaid.}
\label{fig:1}
\end{figure*}

\begin{figure*}
\resizebox{\hsize}{!}{\includegraphics[scale=0.32]{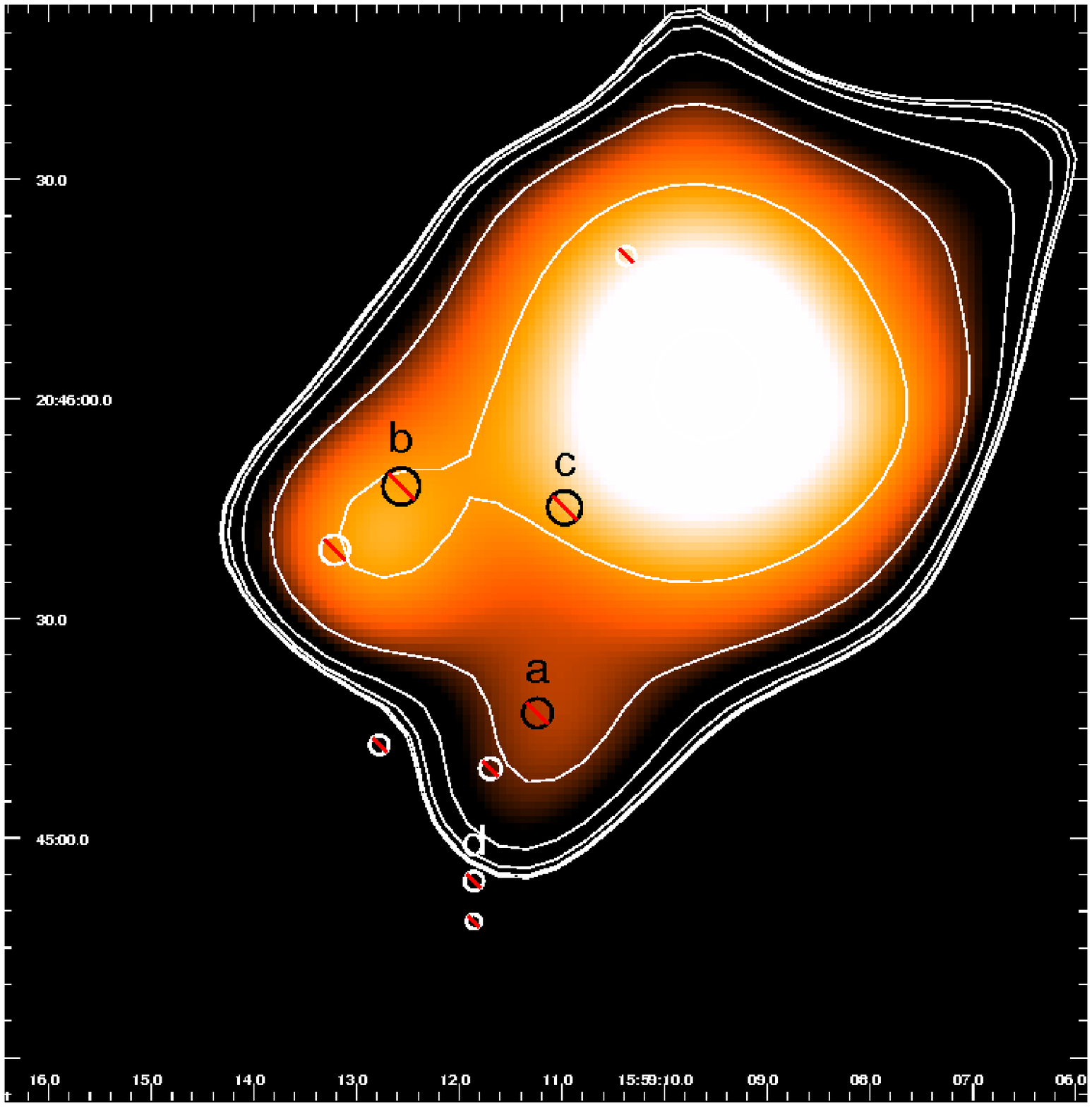}
\includegraphics[scale=0.32]{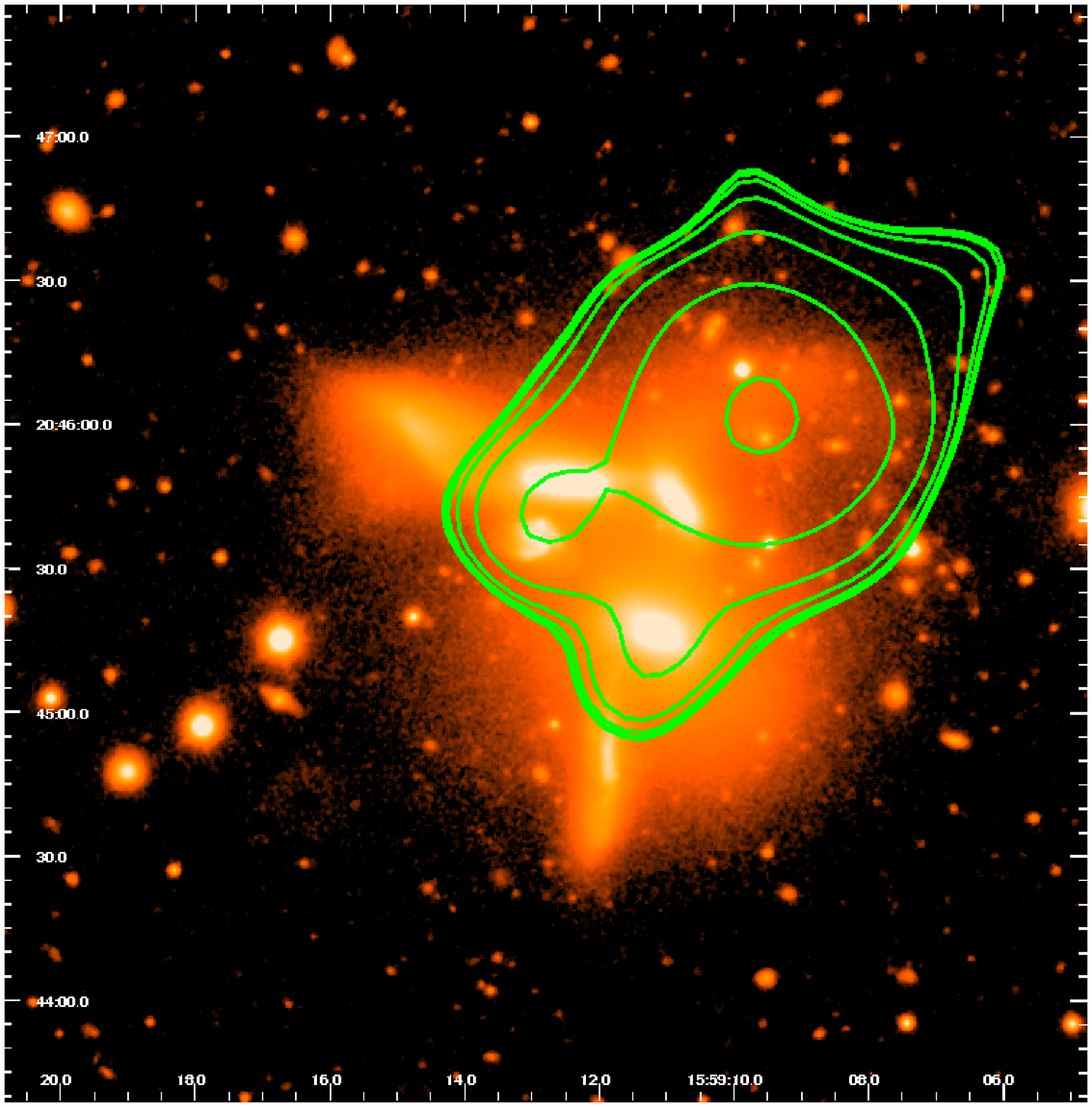}}
\caption{\textit{Left}: raw smoothed X-ray image of the Sextet and its surrounding region with point sources subtracted. \textit{Right}: B+R image of HCG 79 with its optical luminous halo adapted from D08. X-ray contours of the pure diffuse hot emission are superimposed on the optical image; the contours are logarithmically spaced and the first is displayed at $0.21 \times 10^{-5}$ ct s$^{-1}$ pix$^{-1}$.}
\label{fig:2}
\end{figure*}

The X-ray emission in the HCG 79 region was imaged by smoothing the 0.5 to 5 keV \textit{Chandra} data using 
the \texttt{csmooth} routine with a Gaussian filter adopting 3$\sigma$ as the minimum significant signal
within the kernel on an image binned at $\sim$1''/pix. Fig.~\ref{fig:1} shows the resulting image in the region of interest: in the left panel we show the X-ray contours superposed on the optical image and in the right one we show the smoothed X-ray image with optical contours from the HST archive.  X-ray emission can be identified from: a) individual sources connected with the optical galaxies (e.g H79abde, see Table~\ref{tab:100}), b) some diffuse emission between the galaxies in HCG 79, and c) a strong extended source NW of the group and likely associated with a  background cluster of galaxies. Our goal in this study is to properly separate the different components and quantify the emission from the group. 
We therefore produced an X-ray image of the pure extended component, shown in Fig. \ref{fig:2}-left, which we  obtained ``refilling"  all point sources detected (see Section 2.2) with the emission from neighboring regions [CIAO's tool: \texttt{dmfilth}]. The X-ray contours of the diffuse emission were then compared to the optical diffuse halo of HCG 79, shown in  Fig. \ref{fig:2}-right (adapted from D08).

\subsection{Discrete X-ray sources}

\begin{table*}
\caption{Positions, net counts, fluxes and luminosities of the sources detected within and around HCG 79.}
\label{tab:100}
\centering
\begin{tabular}{ccccccl}\hline
\small{Source} & RA(J2000) & Dec(J2000) & Net Counts& $f_{x}$ (0.5-5 keV) & $L_{X}$ & Notes\\
\small{Number} & (hh mm ss) & ($^\circ$ ' '' ) & &\ergcmsec & \ergs&\\ \hline\hline
1 & 15 59 10.974 & +20 45 45.11 & $8.9\pm3.3$ & $6.9\times10^{-16}$ & $3.2\times10^{38}$ & H79c\\
2 & 15 59 11.200 & +20 45 17.07 & $26.0\pm5.2$ & $2.0\times10^{-15}$ & $9.5\times10^{38}$ & H79a\\ 
3 & 15 59 11.856 & +20 44 54.11 & $63.4\pm8.0$ & $4.9\times10^{-15}$ & $2.3\times10^{39}$ & H79d\\
4 & 15 59 11.857 & +20 44 48.58 & $9.4\pm3.3$ & $7.3\times10^{-16}$ & $3.4\times10^{38}$ & H79d\\
5 & 15 59 12.564 & +20 45 48.02 & $38.5\pm6.5$ & $3.0\times10^{-15}$ & $1.4\times10^{39}$ & H79b\\\hline
6 & 15 59 04.430 & +20 45 27.37 & $9.5\pm3.2$ & $7.4\times10^{-16}$ & -- &\\
7 & 15 59 10.345 & +20 46 19.50 & $45.5\pm6.9$ & $3.5\times10^{-15}$ & -- & \tiny{FIRST}\\
8 & 15 59 11.695 & +20 45 09.47 & $9.3\pm3.2$ & $7.2\times10^{-16}$ & -- & \tiny{SDSS }\\
9 & 15 59 12.778 & +20 45 12.70 & $7.3\pm3.0$ & $5.7\times10^{-16}$ & -- &\\
10 & 15 59 13.215 & +20 45 39.34 & $14.6\pm4.0$ & $1.1\times10^{-15}$ & -- & near H79e\\\hline
\end{tabular} 
\begin{list}{}{}
\item[ ]{\textbf{Notes.} Flux $f_x$ in the broad energy band (0.5-5) keV is calculated assuming a fixed power law with index $\Gamma=1.7$ and Galactic line-of-sight absorption. Counts from the detection algorithm.}
\end{list}
\end{table*} 

We employed a source detection algorithm from the CIAO package (\texttt{wavdetect}) on a section of 
the (0.5-5) keV band image that includes HCG 79 and surroundings (a box of size $8'\times8'$). 
We used \textit{ds9} to examine all obvious contaminating sources. Several discrete X-ray sources are found within and near HCG 79 and five of them appear to be connected with the galaxies in the group. Only one discrete source was 
identified in the strong diffuse emission NW of HCG 79. It corresponds to a radio detection 
(see Table~\ref{tab:100}, source \#7) and on the high resolution version of the public HST WFPC2 image 
it is spatially coincident with a pair of apparently interacting galaxies, one of which shows a remarkably blue 
disk or optical jets. It may belong to the background cluster as discussed in \cite{palma} but no redshift is available.
Table 2 lists the discrete sources found together with their optical counterparts within HCG 79 whenever available 
or with other identifications from the literature. We report their positions, net counts and 
uncertainties (as estimated by the detection algorithm) and the (0.5-5.0) keV flux obtained 
assuming a constant conversion factor of $7.8\times10^{-17}$  erg cm$^{-2}$ s$^{-1}$ counts$^{-1}$  
(corresponding to a power law with $\Gamma=1.7$ and Galactic line-of-sight absorption 
$N_{H}=3.8\times10^{20}$ cm$^{-2}$, \citealt{kalberla}). Luminosities are derived 
assuming $D=60.4$ Mpc only for sources associated with HCG 79, since no redshift is available  
for the other sources. Background source \# 10 in Table~\ref{tab:100} is located near H79e but it does not coincide with the galaxy nucleus.
The X-ray sources associated with the galaxies show (0.5-5) keV luminosities in the range of $\sim10^{38}-10^{39}$ \ergs\ , which are typical of bright sources found in normal galaxies (see review by \citealt{fabbiano06}).
H79a and H79b show evidence for nuclear activity at radio, MIR and optical wavelengths with H79a a bone fide AGN and H79b a minor merger (D08); the X-ray sources associated with these two galaxies, even if coincident with the nuclear position, are at best very low luminosity AGN.

\subsection{Searching for diffuse emission in Seyfert's Sextet}
\begin{figure}
  \includegraphics[scale=0.40]{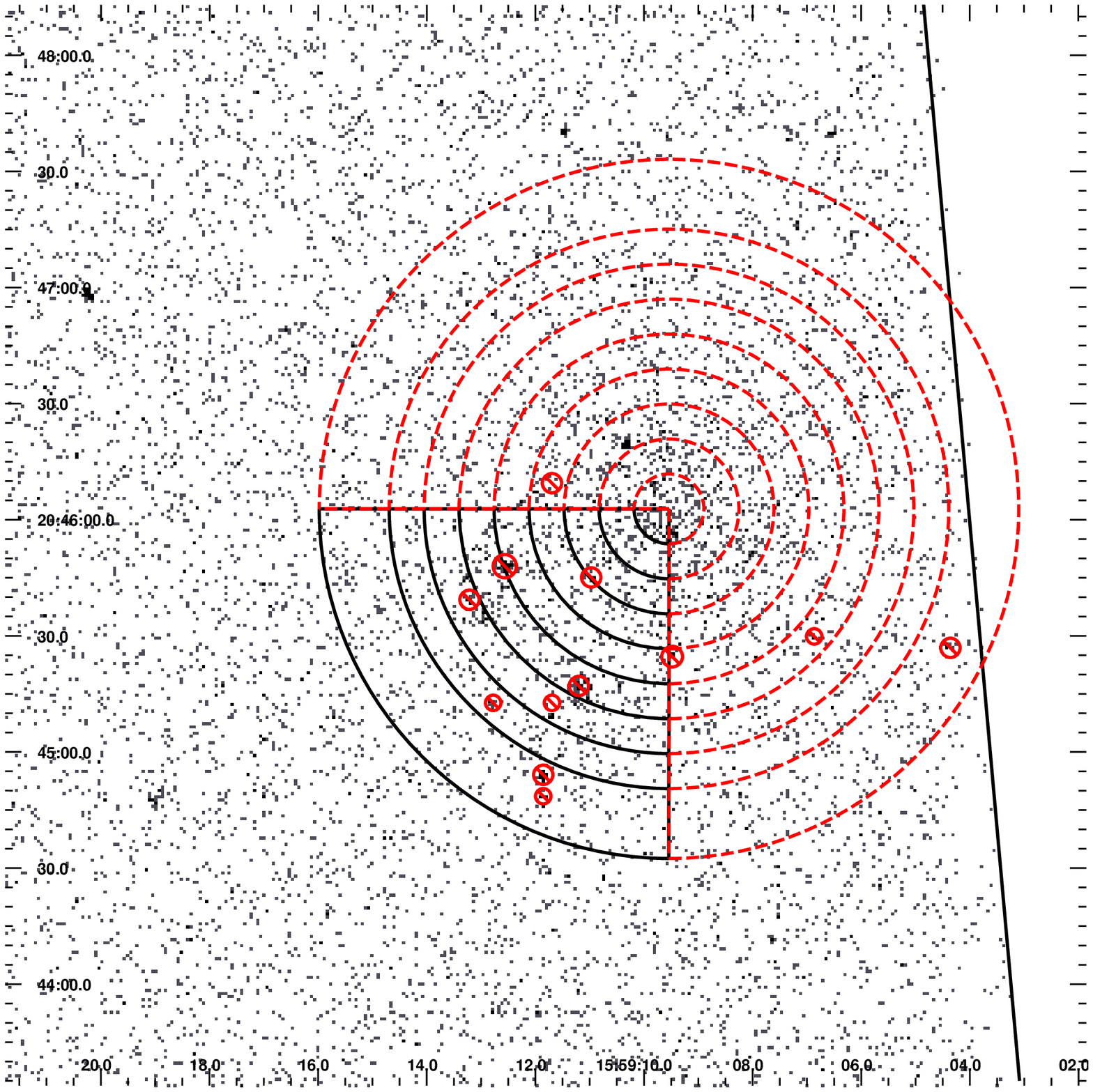}
\caption{\textit{Chandra} image in the (0.5-5) keV energy band. Regions used to derive the radial profile: solid/black for SE region (90 deg to 180 deg counter clockwise from N), dashed/red for the other three quadrants. Black slanted line traces the S3 chip border. Circles indicate detected point-like sources. The color version is available on line.}
  \label{fig:3}
  \end{figure}
\begin{figure*}
\resizebox{\hsize}{!}{\includegraphics[scale=0.32]{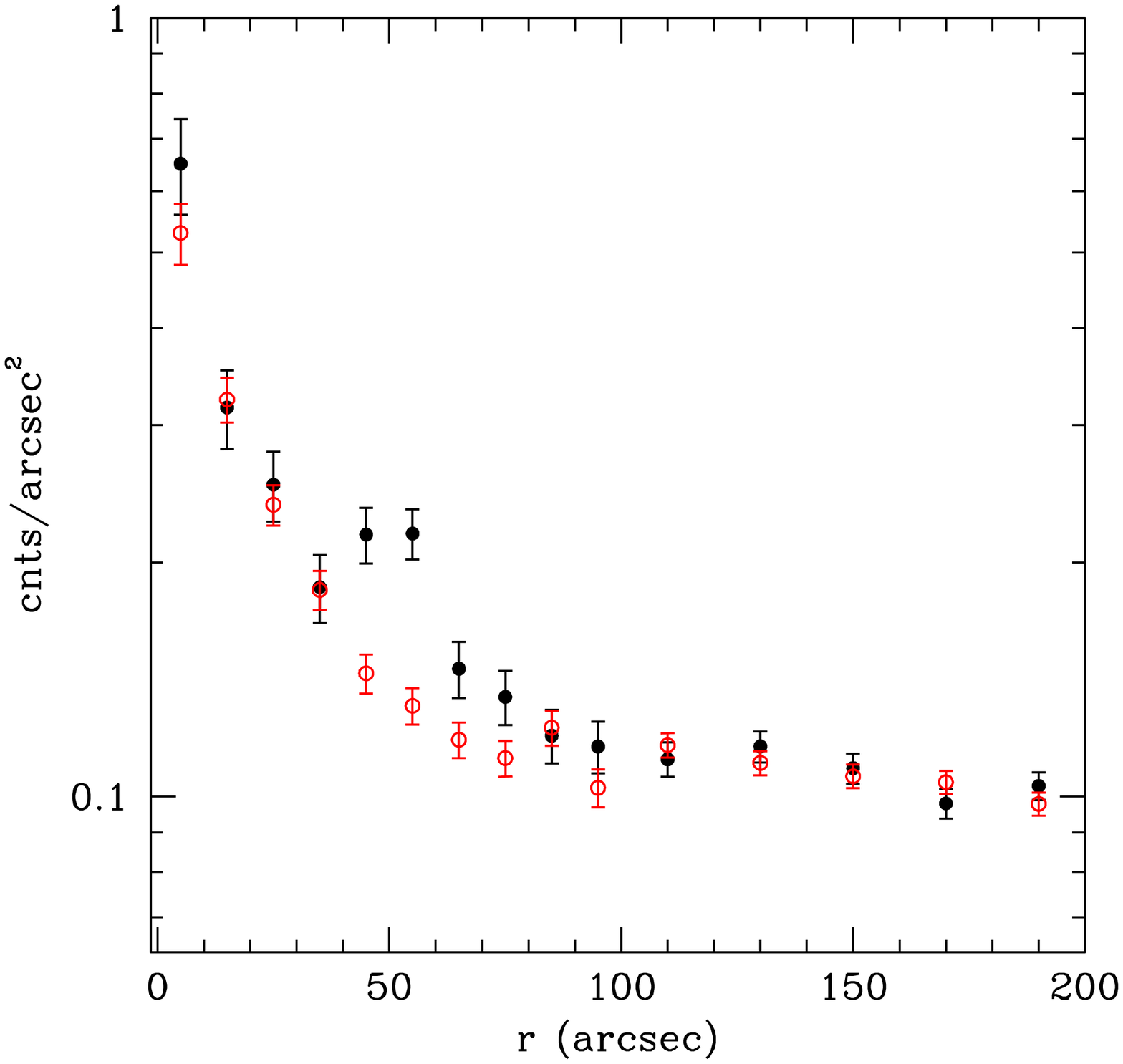}
\includegraphics[scale=0.32]{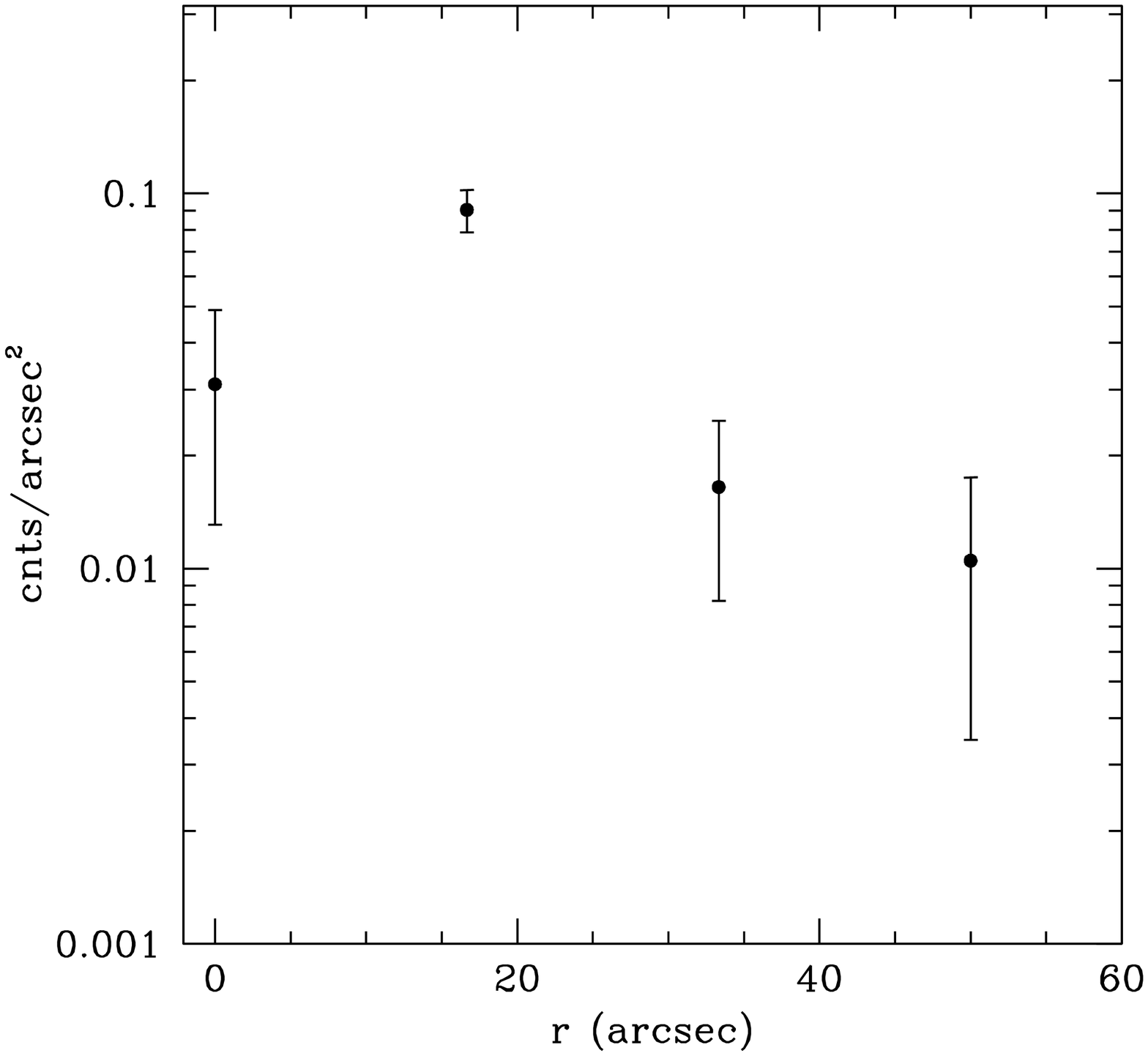}}
\caption{\textit{Left}: radial profiles of the total emission centered on the peak of the cluster's emission.The regions used for the profiles are shown in Fig.\ref{fig:3}. Black/filled symbols refer to the SE quadrant (black solid line in Fig.\ref{fig:3}), red/open symbols refer to the complementary 270 deg region (red dashed line in Fig.\ref{fig:3}). \textit{Right}: radial profile of HCG 79 centered at (RA = $15^{h} 59^{m}{11.9}^{s}$, $\delta=+20^{\circ}45^{'}{31.0}^{''}$) the center of the optical halo (\citealt{darocha05}).} 
\label{fig:4}
\end{figure*}

In order to quantify the amount and distribution of the diffuse emission associated with Seyfert's Sextet we must first estimate 
the contribution from the strong source to the NW of it, which is likely 
related to the cluster of galaxies identified by \cite{palma}. 
The center of this X-ray emission is consistent with a galaxy well resolved by HST which could very well be the cluster cD.
We produced two radial profiles centered on the peak of the cluster emission (RA = $15^{h} 59^{m}{09.54}^{s}$, $\delta=+20^{\circ}46^{'}{02.82}^{''}$), one taken in the SE quadrant (90 deg to 180 deg counter clockwise from N) where HCG 79 is located and a second in the complementary 270 deg azimuthal region contained in the S3 chip, shown in Fig. \ref{fig:3}. Only a small fraction of the outer annuli falls outside the S3 chip, which we have not included in the area considered.
Point sources were removed using circles with radius $1''\div3''$ following source sizes derived by wavdetect. The profiles are plotted in Fig. \ref{fig:4}-left. Their comparison indicates a clear excess of
counts in the SE quadrant particularly in the radial range 40''$\div$80'' which corresponds to the region 
of the group. 
Since the excess emission sits on top of the emission from the cluster which extends out to r$\gtrsim80'' $ (see Fig. \ref{fig:4}-left) we need to include it in our estimate of the background for the group. We have therefore assumed that the emission from the cluster is azimuthally symmetric around its center and estimated the background from the annular region at the same distance from the cluster center as HCG 79 in the azimuthal sectors clockwise E to S (dashed in Fig. \ref{fig:3}). This includes also the instrumental and sky background in the field. 

We tried to obtain a radial profile of the diffuse group emission to better study its spatial distribution, but the presence of the strong source to the NW together with the weakness of the emission from the group prevented us from finding an obvious X-ray peak on which we could center the profile. Therefore we chose to centre the radial distribution on the optical diffuse halo first discussed by \cite{darocha05} (see also Fig. \ref{fig:2}-right). In Fig. \ref{fig:4}-right we show the resulting radial profile. The background for each annulus of the profile was estimated taking into account the emission from the cluster as discussed above. We can see that the profile does not peak at the center of the optical halo but has a stronger emission at a radius of 15''$\div$35''.

Fig. \ref{fig:2}-left shows the most useful representation that we could obtain by generating an X-ray image with 
all discrete sources subtracted. This suggests that the diffuse emission in HCG 79 involves two separate components: 1) the region of galaxy H79a and 2) the region of galaxy H79b and its surroundings, of about 100 counts each. Part of the latter emission might be due to the background source near H79e. This extended emission could be regarded as a halo component specific to those galaxies or the candidate for diffuse X-ray substructure in HCG 79. With the limited statistics and the complexity of the field, it is unlikely that we can resolve the issue at this time. The presence of the background galaxy H79e represents an additional source of noise to our analysis, since in spite of it being further away, it could still contribute: for instance a reasonable luminosity of $\sim10^{40}$ \ergs\  would contribute $\sim 10$ net counts, and a moderate starburst could give a higher contribution.

\begin{figure}
\includegraphics[scale=0.44]{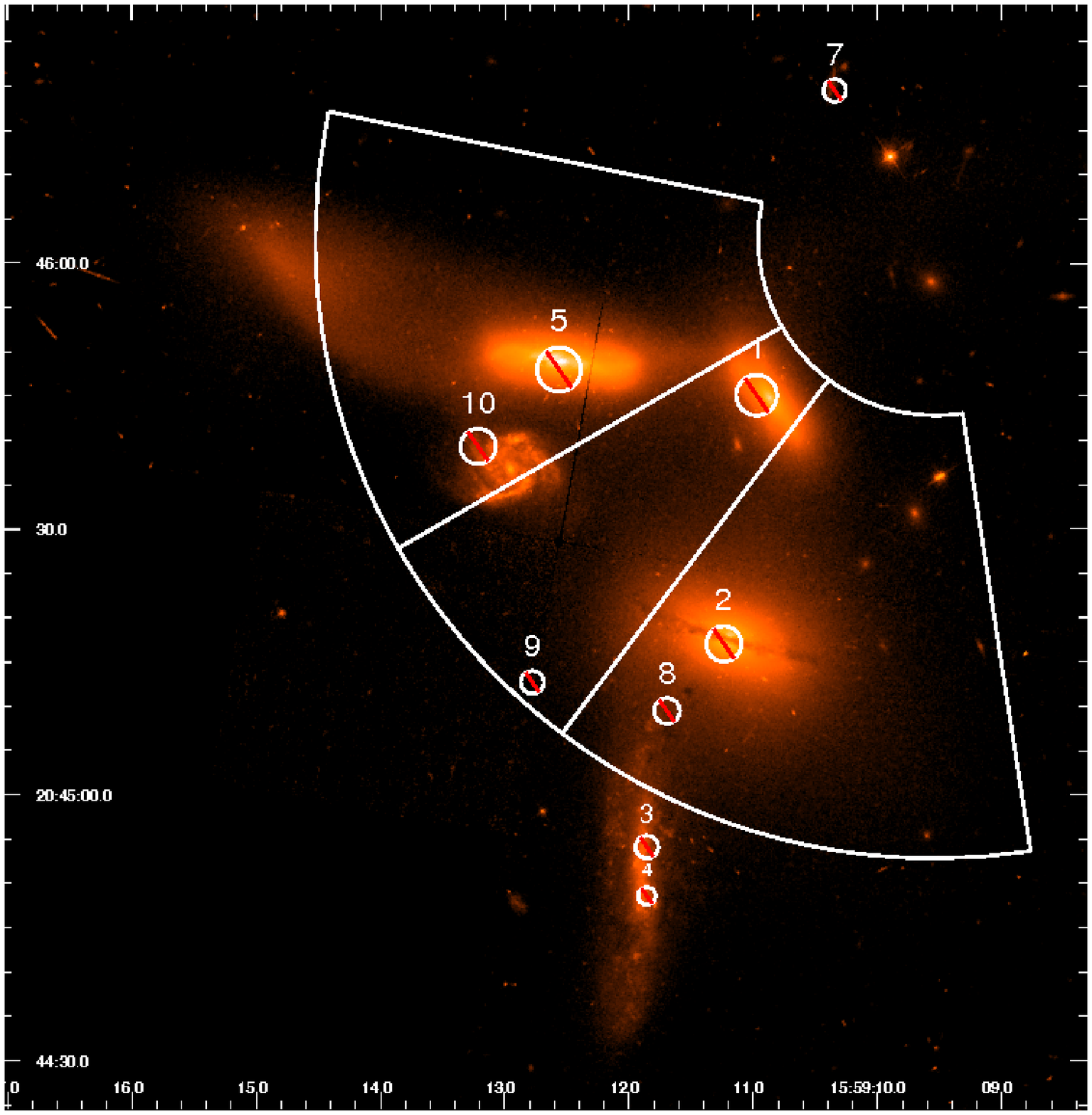}
\caption{Regions used to estimate the emission of HCG 79 relative to its members are shown on the HST image (see text). The first wedge subtends an angle from 78 deg to 120 deg, the second one from 120 deg to 143 deg, the third one from 143 deg to 189 deg counter clockwise from N.}
\label{fig:5}
\end{figure}

In order to better quantify the extended group emission we compared the X-ray surface brightness of 3 regions chosen to separate the galaxy contribution from the intragroup medium (IGM) one (see Fig. \ref{fig:5}). 
The (20''/70'') inner/outer radii are dictated by the evidence of an excess emission and are selected to include almost all of the group galaxies. Point sources were removed using circles with radius $1''\div3''$ and correspond to the sources listed in Tab. \ref{tab:100}. We find marginal evidence that the region between the galaxies  (central sector considered) shows a lower surface brightness than (S$_{B}=0.028\pm0.017\ \rm{cnts/arcsec^2}$ compared to S$_{B}=0.050\pm0.011\ \rm{cnts/arcsec^2}$) confirming the impression given in Fig. \ref{fig:1} and suggesting very little emission in the intergalactic intragroup region.

 \begin{figure*}
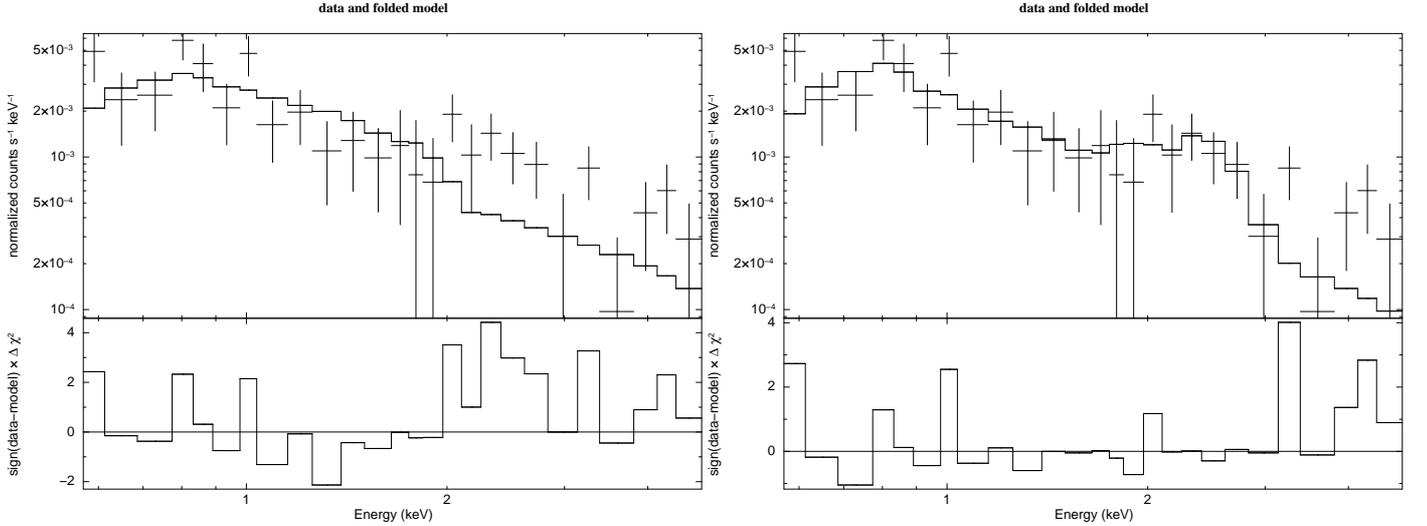

 \resizebox{\hsize}{!}{\includegraphics[angle=-90]{Fig6_left.ps}
 \includegraphics[angle=-90]{Fig6_right.ps}}
\caption{Spectral distribution for the emission compared to spectral models. The spectrum includes the contribution from H79a, H79b and H79d and excludes the other point sources. Spectral models used: \textit{Left}: APEC and Power Law. \textit{Right}: APEC, Power Law and Gaussian at E$\sim 2.4$ keV.}
 \label{fig:6}
 \end{figure*}

\subsection{Spectral Analysis}
We used the \texttt{specextract} script in CIAO, appropriate for extended sources, to create the spectral matrices and data files for the XSPEC analysis. 
In order to improve the signal to noise of our data, we chose to bin the spectra using a minimum of 25 total counts per bin. This is a good compromise between increasing the statistical significance of each spectral bin and having a sufficient number of bins. We used the $\chi^2_{\nu}$ statistics to estimate the goodness of fit.
The Galactic hydrogen column density in the model spectra was fixed at the line of sight value determined from radio surveys ($N_{H}=3.8\times10^{20}$ cm$^{-2}$, \citealt{kalberla}). For the APEC model we used abundances tabulated in \cite{anders}.

We analyzed the spectral data of the Sextet in a slightly smaller region than described above, with inner and outer radii of 40$''$ and 70$''$ to obtain a better S/N (see the region showed in Fig.~\ref{fig:5}); moreover, in order to include H79d as well, we added a box that contains this galaxy rather than expanding the wedge to a larger radius which would add a large portion of the annulus with no emission. We did not try to include H79c due to the high contribution of the cluster in that region. We also considered the counts of the X-ray point sources associated to the optical galaxies H79a, H79b and H79d (which contribute $\sim 137$ net counts), but we have excluded  the point sources unrelated to HCG 79. The background was taken in a region complementary to the region of HCG 79 emission that includes the cluster emission as explained above (see also Fig. ~\ref{fig:3}).
The spectral analysis was based on $316.0\pm23.4$ net counts, consistent with those estimated from the excess in Fig. \ref{fig:4}, given the different extraction regions and the presence of the emission from the point sources.

Although the quality of the data is such that a single power law can reasonably represent the data (albeit a power law with a very steep index $\Gamma=3.0$), we chose to use a combination of a thin plasma and power law to account for the different components expected to be present: a diffuse ISM/IGM and discrete sources/AGN in the galaxies. When fitting the X-ray spectrum of the source, we found that we could not constrain well the photon index of the assumed power-law component. This component however was necessary to account for emission at high energy. Thus we fixed the photon index at 1.7 and we fitted only the temperature and the two normalizations. 
With a fixed abundance at $50\%$ of the solar value, we found an X-ray temperature of kT=(0.3$\pm$0.2) keV and $\chi^2_{\nu}=1.6$ for 23 d.o.f. (Fig. \ref{fig:6}-left).

The high value of $\chi^2_{\nu}$ is primarily due to an excess emission at $E\sim2$ keV, and to a lesser degree to an  excess around 4 keV. We know from the analysis of the \textit{Chandra} background (http://cxc.harvard.edu/contrib/maxim/bg) that there is an Au line contributing at $\sim 2.2$ keV, while no obvious equivalent is visible at 4-5 keV. Before dismissing the $\sim 2$ keV feature as due to a faulty background subtraction, we looked at the 2-3 keV spectral region in different areas of the detector.  We found, for example, that the excess is more prominent in the area around H79a, relative to the area around H79d. We also noticed that different background areas yield different levels of excess emission around $\sim2-2.5$ keV, from no excess to a recognizable bump.  We also looked at the possibility of contribution from one of the detected sources, which we include in the overall spectra.  We concluded that the excess is most likely due to fluctuations in the background and we resolved to model it with a Gaussian line, with peak energy at $\sim 2.4$ keV, which reduces the  $\chi^2_{\nu}$ value to a more reasonable value $\chi^2_{\nu}=1.0$ for 20 d.o.f. (Fig. \ref{fig:6}-right). 


With the inclusion of the Gaussian line to account for the Au line, the best fit temperature is (0.4$\pm0.2$) keV.  The corresponding unabsorbed flux and  luminosity for the plasma component are $f_X\sim 4.3 \times 10^{-15}$ \ergcmsec, $L_X\sim 2.0\times 10^{39}$ \ergs\ . The  power law component contributes  $f_X\sim 2.0\times 10^{-14}$ \ergcmsec and  $L_X\sim 9.6\times 10^{39}$ \ergs\ . This value is entirely consistent with the integrated X-ray luminosity expected from the low mass X-ray binary component 
present in the sum of all galaxies, derived from the L$_X$(LMXB)/L$_{B}$ relation of \cite{kim}.

Using the spectral information obtained from the whole source, we estimated the luminosity that we can attribute to individual galaxy. For each galaxy we assumed an elliptical region with major axis from Table 5 of D08, corresponding to the last concentric isophote in the SDSS \textit{g} band. We chose local background regions near each galaxy, which should include the contributions from the instrument, the cluster and any diffuse emission from the group. The results are reported in the last column of Table \ref{tab:1}, where the L$_X$ corresponds to the total excess above the surrounding emission inclusive of point sources if inside the ellipse. Note that in Table \ref{tab:100}, the luminosity L$_X$ refers only to the point sources found from the detection algorithm. As expected, the sum of the contributions from the galaxies is more than 70\% of the total emission detected.

\section{The background cluster source}
\begin{figure*}
\includegraphics[width=8cm]{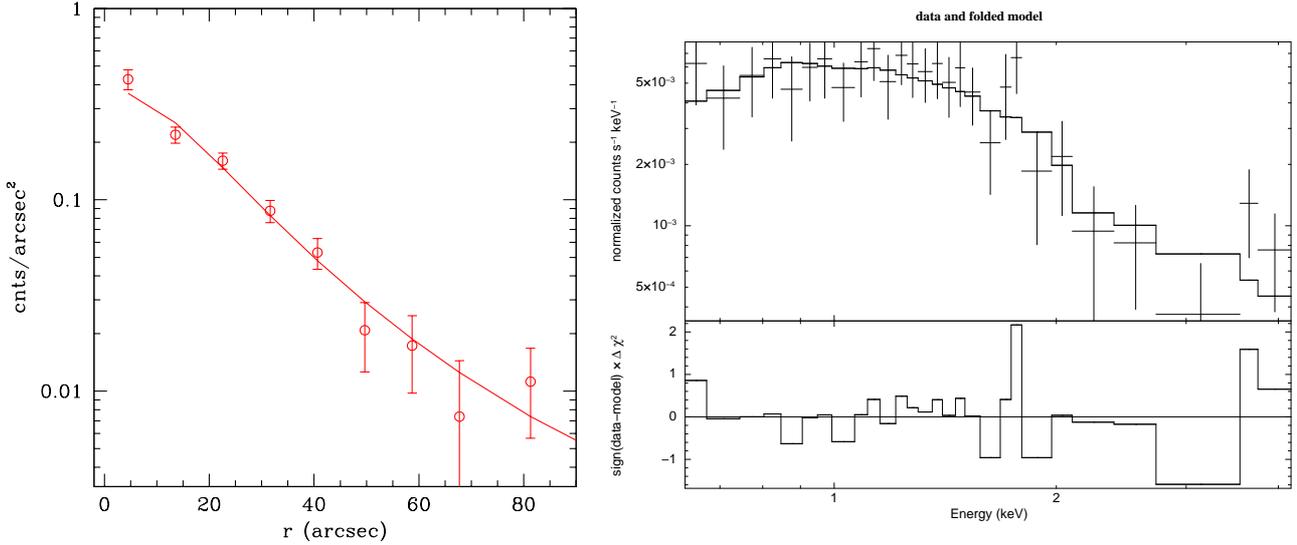}
\vskip -8.0truecm
\hskip 8truecm
\includegraphics[angle=-90,width=9cm]{Fig7_right.ps}
\vskip 0.1truecm
\caption{\textit{Left}: net radial profile centered on the peak of the cluster emission (red open symbols). The solid line represents the best fit obtained with a $\beta$ model with $\beta\sim0.7$ and $r_c\sim25.5''$. \textit{Right}: spectral distribution of the cluster's emission. The spectrum contains about $573.7\pm38.6$ net counts, binned to have at least 18 net counts in each spectral bin. Spectral model is APEC, the fit gives $\chi^2_{\nu}=0.5$ for 26 d.o.f.}
\label{fig:7}
\end{figure*}

In order to better investigate its nature, we have also analyzed the emission from the source to the NW, which we attribute to a background cluster. We have produced a radial profile centered on the NW X-ray peak (Fig. \ref{fig:7}-left). We used concentric annuli with outermost radius of $\sim90''$. We estimated the background in a concentric annulus with inner/outer radii of 100$''$/130$''$ excluding the SE region containing the group. 
We fitted this profile with a $\beta$-model distribution ($S(R)=S_0[1+(R/r_c)^2]^{-3\beta+0.5}$), traditionally used to describe the surface brightness profiles of clusters and groups of galaxies (e.g. \citealt{jones84}), plotted in Fig. \ref{fig:7}-left. The best fit parameters obtained are: $\beta=0.7 \pm0.2$ and $r_c=25 \pm 8$ arcsec, corresponding to a size of 109.31 kpc (if we assume the redshift $z=0.3$ as the lower limit estimated by \citealt{palma}).

The X-ray spectrum for the NW source and relative background were extracted from a circle of 65'' radius centered on the cluster's peak, but excluding a SE box at the position of the group, and the adjacent annulus, with $80''<r<130''$ respectively. The spectrum contains $573.7\pm38.6$ net counts (Fig. \ref{fig:7}-right). To obtain a 
reasonable statistical significance and no negative bins in the energy range considered, we chose to bin the data to have 18 net counts per bin.

We assumed a thin plasma model (APEC) with the abundance fixed at 50\% of the solar value and a Galactic hydrogen column density ($N_{H}=3.8\times10^{20}$ cm$^{-2}$, \citealt{kalberla}). The fit gives $\chi^2_{\nu}=0.5$ for 26 d.o.f..
Given the limited quality of the data, we could only derive a temperature kT$\sim6$ keV ($\gtrsim$3 keV), which corresponds to an  unabsorbed flux $f_X \sim5.52  \times 10^{-14}$ \ergcmsec. Assuming photometric redshift $z=0.3$ (the lower limit reported by \citealt{palma}) and  kT$\sim6$ keV we found a luminosity of $L_X\sim 1.5\times 10^{43}$ \ergs\  in the (0.5-5) keV band.
Assuming that the source is spherically symmetric, we can correct for the missing area due to the presence of the group and estimate that the total luminosity from the cluster amounts to $L_X\sim 2\times 10^{43}$ \ergs\  in the (0.5-5) keV band.
The parameters found are consistent with a low luminosity cluster of galaxies (e.g. \citealt{pacaud07}, \citealt{sarazin86}), but we need more detailed information on this source to draw firm conclusions on its nature.

\section{Discussion and Conclusions}
X-ray observations of HCG 79 show that the field is rather complex with an X-ray bright background galaxy cluster accounting for most of the detected
emission. The excellent quality of the \textit{Chandra} data enables us to separate emission from HCG 79 and the cluster with reasonable confidence.
We find the group to be extremely deficient in X-ray emission: the total (0.5-5) keV luminosity originating from HCG 79 is $L_X\sim 1.2\times10^{40}$ \ergs\ , of which only $L_X\sim 2.0\times 10^{39}$ \ergs\  can be ascribed to a gaseous component. Most of the detected counts appear to be associated with individual galaxies (primarily H79a to the W and H79b and its surroundings to the E, cf. Table \ref{tab:1} and Fig. \ref{fig:2}-left) and therefore it is unlikely that we have detected ``bona-fide'' diffuse group emission. The reality of any diffuse emission of course depends upon our assumption that the background cluster source is relaxed without any subcluster emission on the side towards HCG 79. The discontinuity between the low inferred temperature of the emission from HCG 79 and the hot background cluster source adds support to the plausibility of our assumption.

The weakness of any diffuse emission from the intergalactic medium has significant implications for our understanding of compact group evolution.
Theoretical models predict the coalescence of compact groups into a single object (\citealt{governato}; \citealt{atha}), possibly a fossil elliptical, which should be characterized by significant X-ray emission (Jones et al. 2003). Most current models assume that groups coalesce by major merging. Instead the detailed study of HCG 79 (D08) points toward a slow and quiet dissolution process rather than coalescence by major merging, in line with the model by \cite{atha} which indicates that groups embedded in a massive dark matter halo can persist much longer.

HCG 79 is a very compact group with member galaxies too luminous for their small sizes (D08). These ``compact'' galaxies are embedded in an optical halo that contributes 50$\%$ of the total group light. Based on the inferred mass of this diffuse halo, its color, the isolation of the group, and the size/luminosity similarity of group members to bulges of spirals in the neighborhood, D08 proposed that current members are remnants of larger galaxies embedded in their own debris halo. If this interpretation is correct, then the major challenge connected with this group is to explain the small amount of hot gas that it contains.

\cite{verdes} has argued that HCG 79 is also deficient in cold gas based on VLA observations of the HI content. Subsequent single-dish GBT data (\citealt{borth}) has reduced the evidence of cold gas deficiency in HCG 79 if estimated based upon current galaxy morphologies. The total HI mass (4.3$\times$10$^9$ M$_{\odot}$, \citealt{borth}) associated with the group is typical of a single spiral galaxy, and at least 2$\times$10$^9$ M$_{\odot}$ is coincident with H79d (D08). The low surface brightness component found with single dish observations might be the remnant of an earlier stripping event. The expected HI content in HCG 79 is uncertain, as also suggested by \cite{verdes} and D08, due to the uncertain morphologies of HCG 79 members when they first entered the system. If we assume, as suggested in D08, that at least some of the current early-type members entered the group as spirals then we would infer a much higher gas deficiency. On the contrary HCG 79 is a HI-normal system if the intruders were passive spiral galaxies (galaxies with a spiral morphology and lack of star-formation activity; \citealt{dressler}, \citealt{poggianti}). \cite{bundy} show that these passive galaxies tend to be Sa-Sb types, more abundant at low mass, and presumably gas poor; they might not be as rare as inferred by \cite{goto}.

In order to evaluate the peculiarity of HCG 79 we searched for other compact groups where diffuse optical halos have been measured and for which X-ray data is also available. We found four compact groups: HCG 15, HCG 51, HCG 90, HCG 92, with available X-ray observations. All four are bona fide compact groups possessing n$>$ 4 accordant redshift members and clear signs of dynamical evolution. We use the diffuse optical halo light fraction as a chronometer to rank the groups by age -- groups with a larger halo light fraction should be older.

The first comparison is with the well studied Stephan's Quintet (HCG 92); both typify evolved compact groups. Their significant diffuse light fractions leave little doubt that they are physical aggregates with Stephan's Quintet younger (and more massive) than Seyfert's Sextet.
\object{HCG 92} is less relaxed and contains a lower diffuse light fraction ($\sim30$\%) (\citealt{moles}). Two spirals in Stephan's Quintet are in the process of transformation from spiral into lenticular galaxies and the presence of a high velocity intruder produces a large scale shock confirmed by \textit{Chandra} and XMM-Newton observations (\citealt{trinchieri03,trinchieri05}). Separating shocked and diffuse X-ray components enabled \cite{trinchieri05} to estimate the luminosity of the diffuse (non-shocked) X-ray component in HCG 92: $L_X\sim5\times10^{41}$ \ergs\  (0.5-2 keV; \citealt{trinchieri05}). Although not particularly X-ray luminous, this value is two orders of magnitude larger than the extremely low X-ray luminosity of HCG 79.

Another group of possible relevance involves \object{HCG 90} which consists of a core of three bright galaxies (two ellipticals and a disturbed disk galaxy) surrounded by an extended loose group. The contribution of the diffuse light to the total light is of the order of $\sim 45 \%$ L$_{tot}$, close to HCG 92 and 79. Low luminosity diffuse X-ray emission was reported by \cite{white} who discussed the observed lack of hot gas in terms of expectations from the dynamical state of the system and the evidence for a large reservoir of diffuse optical light. It is interesting to note that, in analogy to what is observed in HCG 79, the X-ray emission in HCG 90 is predominantly concentrated on the galaxies of the group (see Fig. 10 of \citealt{white}). Unfortunately the analogy breaks down there because the luminosity of the gaseous component in HCG 90, $L_X\sim1.6\times10^{41}$ \ergs\  is two orders of magnitude higher than in HCG 79.

The other two groups considered, \object{HCG 15} and \object{HCG 51}, show smaller percentages of diffuse light,  $\sim 20$\% and $\sim30$\% respectively (\citealt{darocha08}). These smaller values would point to dynamically younger systems and hence to smaller fractions of stripped gas. In both cases the intragroup medium content is significantly higher than in HCG 79, L$_X\sim3.2\times10^{41}$ \ergs\  for HCG 15 (\citealt{rasmussen}) and L$_X\sim3.0\times10^{42}$ \ergs\  for HCG 51 (\citealt{sun}).

All the above examples involve systems more massive than HCG 79. Although the velocity dispersion based on a small number of galaxies is a poor indicator of the size of the group potential well, the total luminosities (in stars) and the total masses inferred for these systems are 1-2 order of magnitude larger than estimated for HCG 79 (M$_{tot}\sim3\times10^{11}$ M$_{\odot}$, \citealt{nishiura}). If we then consider that a significantly smaller mass system might not be able to produce enough hot gas and retain it, we could separate issues and consider that the deficit of hot gas in HCG 79
might not be related to its evolution, but simply to its shallow potential and low mass. Had it been possible to have a reliable measure of the total mass of the system we would have tested this hypothesis. At the present time however we notice that the total optical luminosity, HI mass and total X-ray luminosity (in gas and in ``binaries'') are all consistent with what is expected from a normal galaxy. If the final fate of compact groups is in the formation of a single galaxy, then Seyfert's Sextet is likely to become a relatively isolated, normal [rather than giant] galaxy.

\begin{acknowledgements}
S.T., G.T. and A.W. acknowledge financial support from the Italian Space Agency ASI under contract ASI-INAF agreement I/009/10/0. M.R. acknowledges financial support from contract P-82389 from CONACYT and IN102309 from DGAPA-UNAM.

\end{acknowledgements}

\clearpage
\onecolumn

\listofobjects

\begin{thebibliography}{}

\bibitem[Anders \& Grevesse (1989)]{anders}
Anders, E., Grevesse, N. 1989, GeCoA, 53, 197

\bibitem[Athanassoula et al. (1997)]{atha}
Athanassoula, E., Makino, J., Bosma, A. 1997, MNRAS, 286, 825


\bibitem[Borthakur et al. (2010)]{borth}
Borthakur, S., Yun, M. S., Verdes-Montenegro, L. 2010, ApJ, 710, 385

\bibitem[Bundy et al. (2010)]{bundy}
Bundy, K., Scarlata, C., Carollo, C. M., et al. 2010, ApJ, 719, 1969


\bibitem[Da Rocha et al. (2005)]{darocha05}
Da Rocha, C., Mendes de Oliveira, C. 2005, MNRAS, 364, 1069

\bibitem[Da Rocha et al. (2008)]{darocha08}
Da Rocha, C., Ziegler, B. L., Mendes de Oliveira, C. 2008, MNRAS, 388, 1443

\bibitem[Dressler et al. (1999)]{dressler}
Dressler, A., Smail, I., Poggianti, B. M., et al. 1999, ApJS, 122, 51

\bibitem[Durbala et al. (2008)]{durbala}
Durbala, A., del Olmo, A., Yun, M. S., et al. 2008, AJ, 135, 130


\bibitem[Fabbiano (2006)]{fabbiano06}
Fabbiano, G. 2006, ARA\&A, 44, 323

\bibitem[Finoguenov et al. (2007)]{finoguenov}
Finoguenov, A., Ponman, T. J., Osmond, J. P. F. \& Zimer, M. 2007, MNRAS, 374, 737

\bibitem[Goto et al. (2003)]{goto}
Goto, T., Okamura, S., Sekiguchi, M., et al. 2003, PASJ, 55, 757

\bibitem[Governato et al. (1991)]{governato}	
Governato, F., Bhatia, R., Chincarini, G. 1991, ApJ, 371L, 15


\bibitem[Jones \& Forman (1984)]{jones84}
Jones, C., Forman, W. 1984, ApJ, 276, 38

\bibitem[Jones et al.(2003)]{jones03}
Jones L.R., Ponman T.J., Horton A., et al. 2003, MNRAS, 343, 627

\bibitem[Kalberla et al.(2005)]{kalberla}
Kalberla, P.M.W., Burton, W.B., Hartmann, D., et al. 2005, A\&A, 440, 775 

\bibitem[Kim \& Fabbiano (2004)]{kim}
Kim, D.-W., \& Fabbiano, G. 2004, ApJ, 611, 846

\bibitem[Hickson (1982)]{hickson82}
Hickson, P. 1982, ApJ, 255, 382


\bibitem[Iovino (2002)]{iovino}
Iovino, A. 2002, AJ, 124, 2471


\bibitem[Moles et al. (1998)]{moles}
Moles, M., Marquez, I., Sulentic, J. W. 1998, A\&A, 334, 473

\bibitem[Morita et al. (2006)]{morita}
Morita, U., Ishisaki, Y., Yamasaki, N.Y., et al. 2006, PASJ, 58, 719

\bibitem[Mulchaey (2000)]{Mulchaey}
Mulchaey, J.S., 2000, ARA\&A, 38, 289



\bibitem[Nishiura et al. (2000)]{nishiura}
Nishiura S, Murayama, T., Shimada, M., et al. 2000, AJ, 120, 2355


\bibitem[Pacaud et al. (2007)]{pacaud07}
Pacaud, F., Pierre, M., Adami, C., et al. 2007, MNRAS, 382, 1289

\bibitem[Palma et al.(2002)]{palma}
Palma, C., Zonak, S. G., Hunsgberger S. D., et al. 2002, AJ 124, 2425

\bibitem[Pildis et al. (1995)]{pildis}
Pildis, Rachel A., Bregman, J. N., Evrard, A. E. 1995, ApJ, 443, 514

\bibitem[Poggianti et al. (1999)]{poggianti}
Poggianti, B. M., Smail, I., Dressler, A., et al. 1999, ApJ, 518, 576


\bibitem[Rasmussen et al. (2008)]{rasmussen}
Rasmussen, J., Ponman, T. J., Verdes-Montenegro, L., et al. 2008, MNRAS, 388, 1245

\bibitem[Sarazin (1986)]{sarazin86}
Sarazin C. L. 1986, RvMP, 58, 1

\bibitem[Sulentic (1987)]{sulentic87}
Sulentic, Jack W. 1987, ApJ, 322, 605



\bibitem[Sulentic et al. (2001)]{sulentic01}
Sulentic, J. W., Rosado, M., Dultzin-Hacyan, D., et al. 2001, AJ, 122, 2993

\bibitem[Sun et al. (2009)]{sun}
Sun, M., Voit, G. M., Donahue, M., et al. 2009, ApJ, 693, 1142



\bibitem[Trinchieri et al. (2003)]{trinchieri03}
Trinchieri, G., Sulentic, J., Breitschwerdt, D., Pietsch, W. 2003, A\&A, 401, 173

\bibitem[Trinchieri et al. (2005)]{trinchieri05}
Trinchieri, G., Sulentic, J., Pietsch, W., Breitschwerdt, D., 2005, A\&A, 444, 697



\bibitem[Verdes-Montenegro et al. (2001)]{verdes}
Verdes-Montenegro, L., Yun, M. S., Williams, B. A., et al. J. 2001, A\&A, 377, 812

\bibitem[White et al. (2003)]{white}
White, P. M., Bothun, G., Guerrero, M. A., et al. 2003, ApJ, 585, 739


\end{thebibliography}
\end{document}